\documentclass[a4paper,11pt]{article}
\usepackage[utf8]{inputenc}
\usepackage{jinstpub} 
\usepackage{lineno}
\usepackage{textcomp}
\usepackage{mathcomp}
\usepackage{subcaption}
\usepackage{mathtools}
\usepackage{graphicx}
\usepackage{grffile}

\title{\boldmath Time performance of USTC-IME LGAD under synchrotron light source focused X-ray}

\author[a]{T.A. Wang,}
\author[a]{D. Zhang,}
\author[a]{Z. Liang,}
\author[b]{C.Z. Li,}
\author[c]{J.Q. Luo,}
\author[a]{Y. Nie,}
\author[a]{T.S. Yue,}
\author[a]{Z.B. Tang,}
\author[a]{H. Liang,}
\author[b]{T.P. Yu,}
\author[c]{W.B. He,}
\author[a,1]{Y.W. Liu\note{Corresponding author.}}
\affiliation[a]{Department of Modern Physics, University of Science and Technology of China,\\
Hefei 230026, China}
\affiliation[b]{College of science, National University of Defense Technology,\\
Changsha 410073, China}
\affiliation[c]{Key Laboratory of Nuclear Physics and Ion-beam Application (MOE), Institute of Modern Physics, Fudan University,\\
Shanghai 200433, China}

\emailAdd{yanwen@ustc.edu.cn}

\abstract{The time performance of Low Gain Avalanche Diodes (LGADs), designed by the University of Science and Technology of China (USTC) and fabricated by the Institute of Microelectronics of Chinese Academy of Sciences (IME), was characterized at the Shanghai Synchrotron Radiation Facility (SSRF). The experiment was conducted at the BL16B1 beamline, which delivers a focused X-ray beam with a diameter of 500 \textmu\text{m}, a repetition period of 2 ns, and a photon energy of 10 keV. Using a fast oscilloscope, waveforms containing consecutive signal pulses were recorded within a 50 ns time window. The LGAD under test successfully resolved the period of the SSRF. To reduce the dependence on the number of overlapping pulses in a waveform, a global template fit was employed to extract pulse-by-pulse information from the acquired waveforms. The time resolution was then estimated using a combined profile likelihood approach, with a restricted maximum-likelihood (REML) correction accounting for the per-waveform degrees of freedom, giving $136.6 \pm 0.1$ ps. The analysis chain was then calibrated with the control samples, recorded with different aluminum shields and almost free of pile-up, where the relative time of arrival (TOA) method based on constant-fraction discrimination (CFD) was taken as the reference to extract the time resolution. The corrected time resolution is $119.7 \pm 0.1$ (stat.) $\pm 2.6$ (syst.) ps for the LGAD. The effect of random photon absorption depth on the time resolution of LGADs was studied through dedicated simulations.}

\keywords{Particle tracking detectors (Solid-state detectors); Timing detectors; X-ray detectors}

\arxivnumber{2609.00972} 

\begin{document}
\maketitle
\flushbottom

\section{Introduction}
\label{Introduction}

The Low Gain Avalanche Diode (LGAD) is a type of silicon sensor characterized by excellent time resolution, fast rising time, short dead time, and high radiation tolerance~\cite{LGAD1,LGAD2}. In recent years, it has received increasing attention in the field of experimental particle physics. Both the ATLAS and CMS experiments at the Large Hadron Collider (LHC) have adopted LGADs for their upgrade projects~\cite{ATLAS,CMS}. Designed by the University of Science and Technology of China (USTC) and fabricated by the Institute of Microelectronics of Chinese Academy of Sciences (IME), the USTC-IME LGADs have been developed for the ATLAS High Granularity Timing Detector (HGTD) project~\cite{USTC1,USTC2}. Building on this success, the team is exploring additional scientific and technological applications for these sensors.

Owing to a highly doped gain layer, located a few micrometers beneath the charge collection electrode, LGADs provide inherent gain. This enables X-rays to generate sufficient charge within a thin silicon layer for readout by front-end electronics. To investigate this potential, several research groups have begun studying the response of LGADs to X-rays~\cite{Xray1,Xray2,Xray3,Xray4}. LGAD sensors from Hamamatsu Photonics (HPK), Brookhaven National Laboratory (BNL), and Fondazione Bruno Kessler (FBK) were tested at the Stanford Synchrotron Radiation Lightsource (SSRL). The energy resolution and timing performance of these LGADs were characterized as functions of bias voltage and incident photon energy. All tested LGAD sensors successfully resolved the 2.1 ns repetition period of the SSRL. For 35 keV X-ray beams, the HPK 3.1 LGAD at a bias voltage of 150 V and the BNL LGAD with a thickness of 20 \textmu\text{m} at a bias voltage of 50 V achieved the best energy resolution of 6\%. The BNL LGAD showed the best time resolution of 65 ps at a bias voltage of 100~V~\cite{Xray2}. In this work, the USTC team also characterized the time performance of USTC-IME LGADs for hard X-rays at the Shanghai Synchrotron Radiation Facility (SSRF)~\cite{SSRF1}. The SSRF has a single-pulse jitter of 28.3 ps and a repetition period of 2 ns~\cite{SSRF2}. The tested sensors resolved the repetition period. Moreover, the jitter of the source is sufficiently small to allow for an accurate evaluation of the sensor time resolution. Earlier studies applied a smoothing method to correct the baseline fluctuation in waveforms, but analyzed only signals that were well separated in time~\cite{Xray2,Xray3,Xray4}. In this work, a global template fit was employed to extract information on all signal pulses including those affected by preceding pulses, referred to as pile-up effects hereafter.

The tested devices are described in section~\ref{Devices} and detailed information regarding the beamline station and the experimental setup is presented in section~\ref{Experimental setup}. The analysis consists of two steps, followed by a calibration of the analysis chain on control samples, to characterize the time performance. Section~\ref{Preliminary analysis} presents the relative time of arrival (TOA) obtained with constant-fraction discrimination (CFD) and the extraction of the repetition period of the SSRF. Section~\ref{global fit} describes the fine analysis of the time resolution using the global template fit and a combined profile likelihood approach. Section~\ref{chain validation} establishes the systematic uncertainty of the analysis chain, and the corresponding correction to the result, using control samples. To better understand the time resolution, a simulation combining Allpix Squared and TCAD Sentaurus was performed to investigate the signal generation process of X-rays in LGADs~\cite{TCAD,Allpix}, as detailed in section~\ref{simulation section}.

\section{Devices tested}
\label{Devices}

\begin{figure}[htbp]
\centering
\includegraphics[width=.6\textwidth]{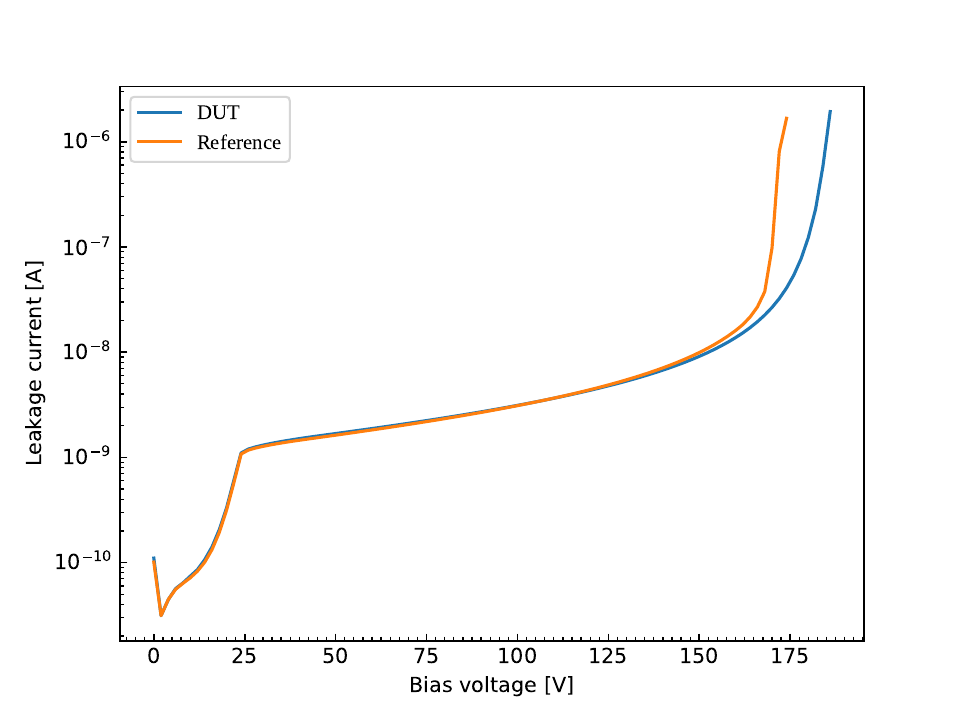}
\caption{I-V curves of the tested sensors.\label{IV}}
\end{figure}

\begin{figure}[htbp]
    \centering
    \begin{subfigure}[b]{0.45\textwidth}
        \includegraphics[width=\textwidth]{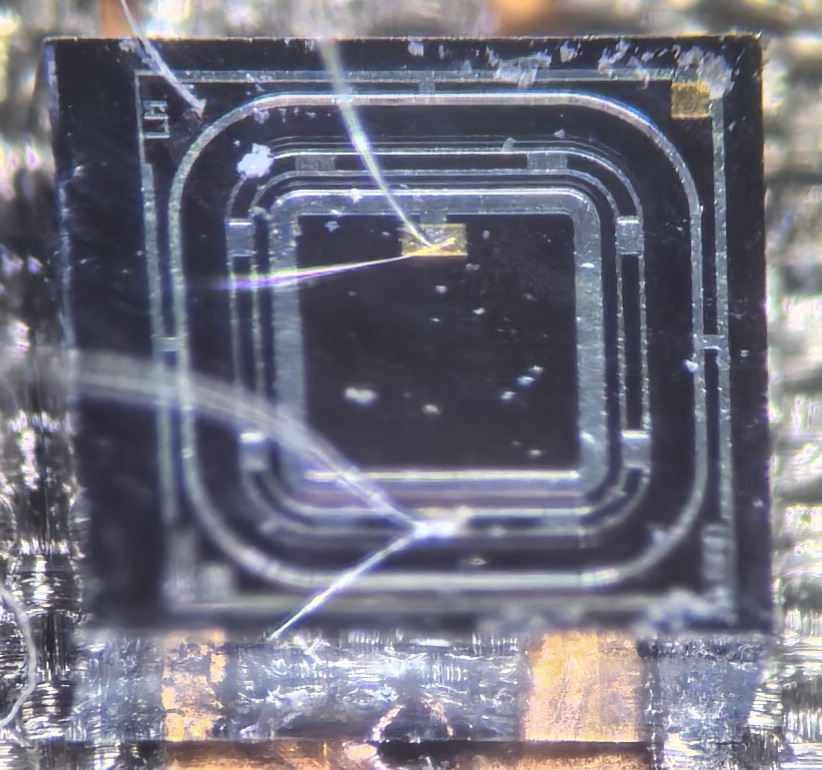}
        \caption{}
        \label{sensor}
    \end{subfigure}
    \hfill
    \begin{subfigure}[b]{0.47\textwidth}
        \includegraphics[width=\textwidth]{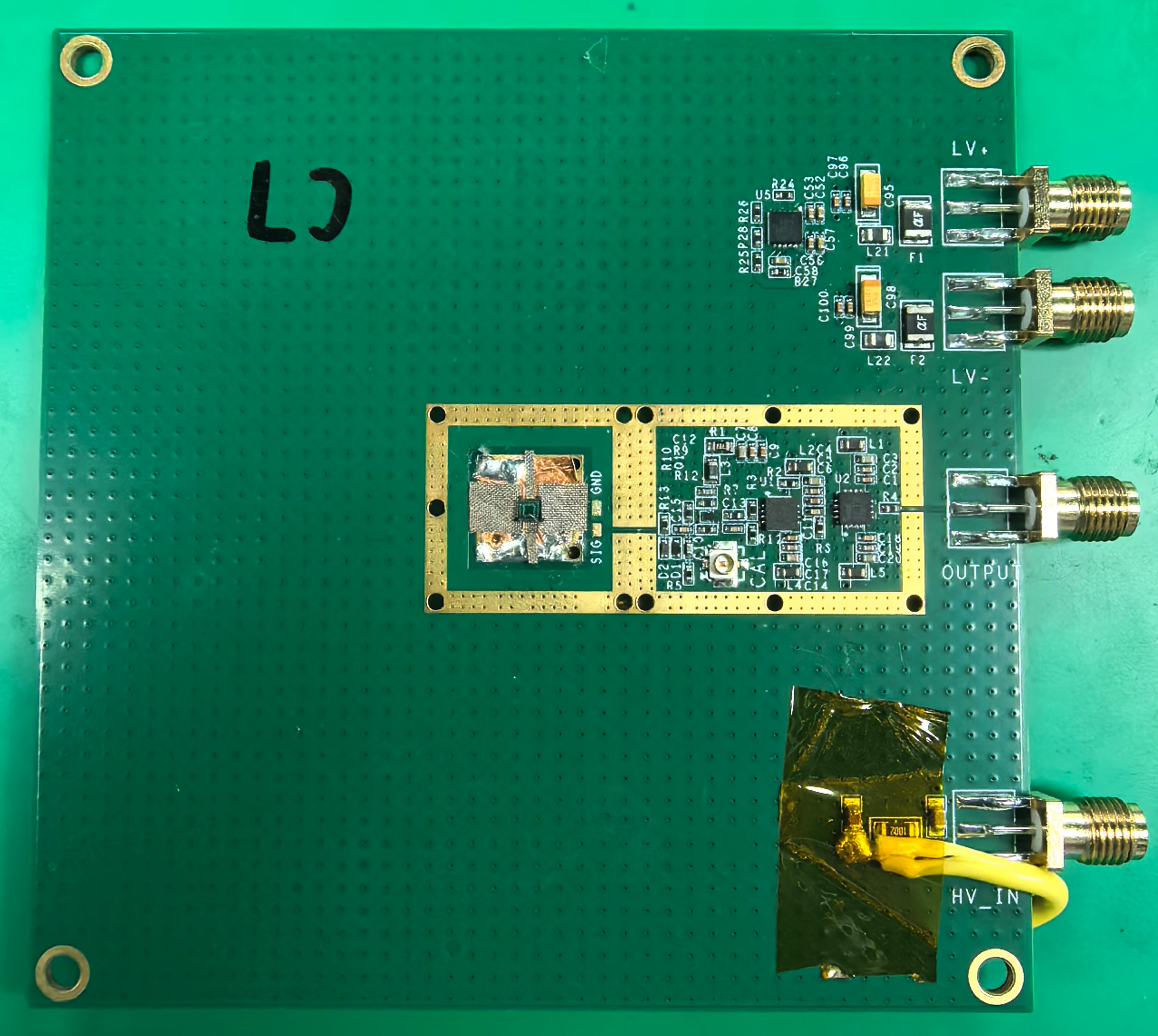}
        \caption{}
        \label{board}
    \end{subfigure}
    \caption{(a) Photograph of a USTC-IME version 2.1 LGAD sensor, showing a window on the surface of the sensitive area. (b) Photograph of the ultra-fast low-noise preamplifier board.}
    \label{sensor and board}
\end{figure}

Two single-pad sensors from the USTC-IME version 2.1 batch were tested~\cite{XYANG}, each with an active thickness of 50 \textmu\text{m}. To increase the detection efficiency, most of the aluminum beneath the passivation layer was removed, creating a large entrance window on the active region surface. Characterized with a beta source at room temperature (26 $\tccentigrade$), both sensors exhibited time resolutions of 39.6 ps. One sensor was used as the device under test (DUT), while the other served as the trigger (referred to as the reference). The two sensors are identical in design, coming from the same version 2.1 production batch; the difference between their I-V curves, shown in figure~\ref{IV}, is attributable to the typical die-to-die variation of the gain-layer doping and of surface currents. Each sensor was wire-bonded to an ultra-fast low-noise preamplifier board for readout~\cite{USTC3}. Because the amplifier board introduces additional impedance, the effective bias voltage applied to the sensor is lower than the voltage applied to the board. In this study, the applied voltage was set to 180 V for both boards, corresponding to a total current of 100 nA, which, based on prior experience, indicates that the sensors were operating under nominal conditions. Photographs of the sensor and the board are shown in figure~\ref{sensor and board}.

\section{Experimental setup at SSRF}
\label{Experimental setup}

The experiment was conducted at the BL16B1 beamline station of the SSRF~\cite{SSRF3}. The beamline delivers a fixed photon energy of 10 keV, and each spill contains approximately 400 photons. The beam is focused to a spot size of about 500 \textmu\text{m} in diameter, which is comparable to the sensitive region of the tested sensors. However, the beamline does not provide a synchronous clock signal, so a direct timing reference is unavailable and the DUT therefore had to be triggered by the reference sensor.

\begin{figure}[htbp]
    \centering
    \begin{subfigure}[b]{0.45\textwidth}
        \includegraphics[width=\textwidth]{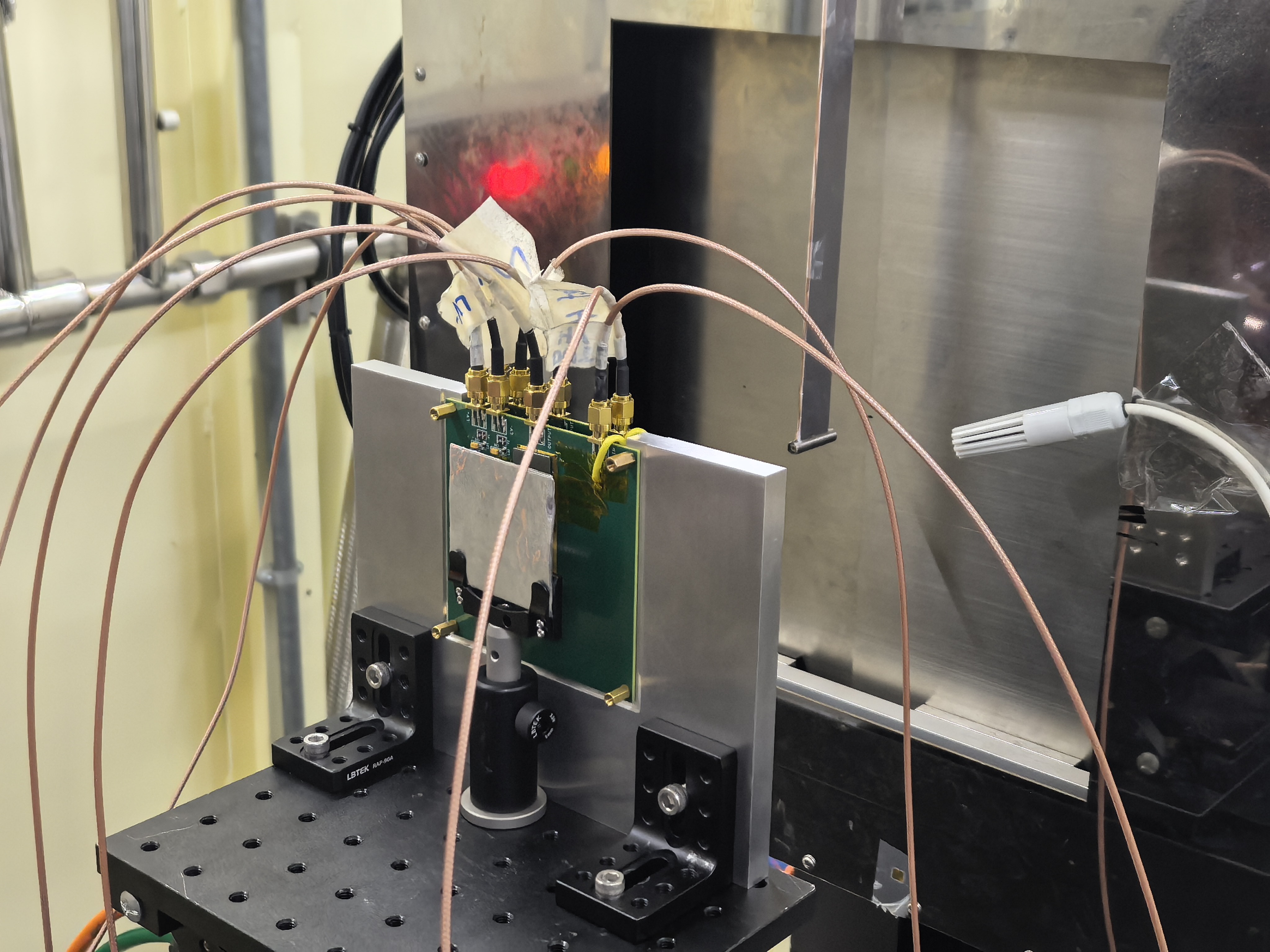}
        \caption{}
        \label{setup1}
    \end{subfigure}
    \hfill
    \begin{subfigure}[b]{0.45\textwidth}
        \includegraphics[width=\textwidth]{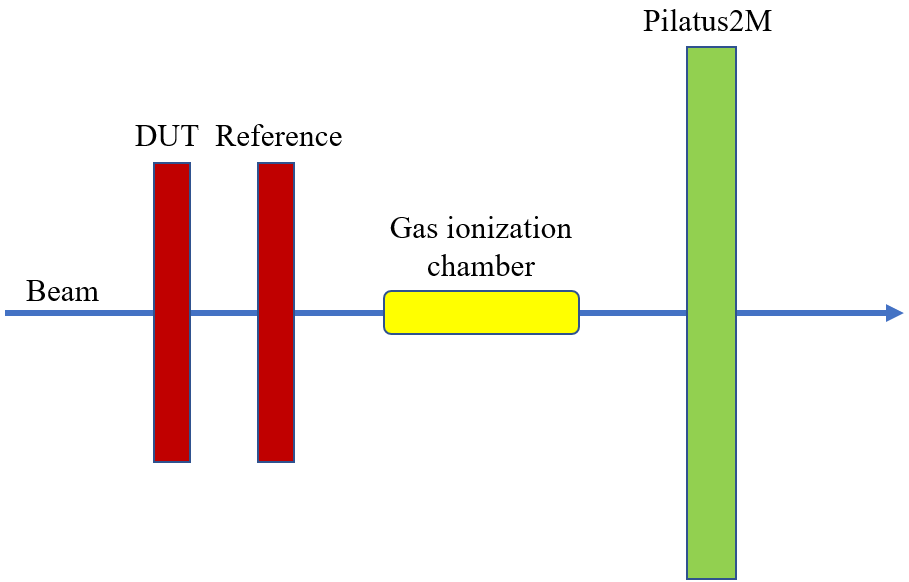}
        \caption{}
        \label{setup2}
    \end{subfigure}
    \caption{Experimental setup at the BL16B1 beamline station of the SSRF. The beam is injected from the left, with the DUT positioned in front of the reference.}
    \label{setup}
\end{figure}

The experimental setup is shown in figure~\ref{setup}. The DUT and the reference sensors, along with their preamplifier boards, were mounted on an optical motion stage capable of fine alignment in three axes. Behind the fixture, a gas ionization chamber and a Pilatus2M detector, provided by the beamline station, were used to measure the beam intensity. An aluminum shield was placed in front of the sensors to attenuate the photon beam. Signals were recorded using an oscilloscope with a bandwidth of 4 GHz and a sampling rate of 40 GS/s. Upon each trigger initiated by the reference sensor, the oscilloscope digitized and recorded the data for a duration of 50 ns. Two Keithley 2470 high-voltage sourcemeters separately supplied the bias voltages for the DUT and the reference, while one Keithley 2231A-30-3 triple-channel DC power supply provided low voltage to power the boards. The oscilloscope and the sourcemeters were controlled by a laptop in the control room.

\begin{figure}[htbp]
    \centering
    \begin{subfigure}[b]{0.45\textwidth}
        \includegraphics[width=\textwidth]{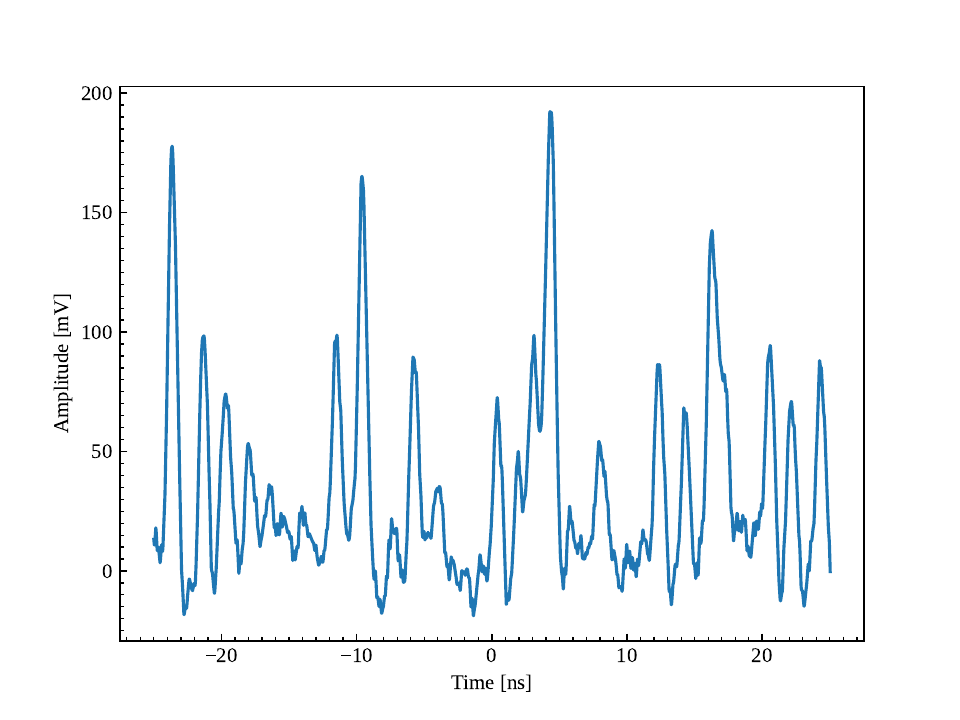}
        \caption{}
        \label{1 mm}
    \end{subfigure}
    \hfill
    \begin{subfigure}[b]{0.45\textwidth}
        \includegraphics[width=\textwidth]{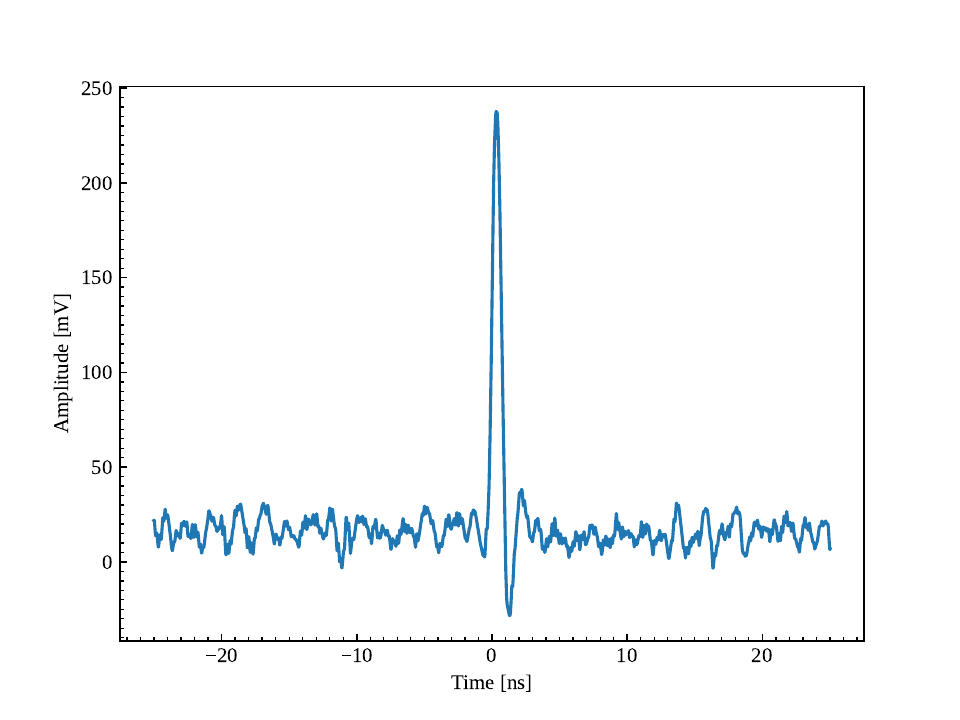}
        \caption{}
        \label{2.2 mm}
    \end{subfigure}
    \caption{(a) Representative waveform with a 1-mm-thick shield. (b) Representative waveform with a 2.2-mm-thick shield.}
    \label{waveforms}
\end{figure}

The photon energy of 10 keV is very close to the most probable energy deposited by a minimum ionizing particle (MIP) in a 50-\textmu\text{m} silicon layer. We aimed to ensure that only one photon could be detected per spill, in order to study the difference in time performance of USTC-IME LGADs when detecting X-rays and MIPs. After numerous attempts, according to the average signal amplitude and the trigger rate, appropriate waveforms were observed with a 1-mm-thick aluminum shield. Figure~\ref{1 mm} shows a representative waveform in this situation. With a thicker shield, the average signal amplitude remained unchanged, while both the number of pulses in a waveform and the trigger rate decreased, indicating that no more than one photon could pass through the shield per beam spill. Therefore, the shield thickness of 1 mm was adopted for most of the experiment to acquire the primary data set. Additional data sets were also collected using thicker shields. For example, waveforms containing signals that were well separated in time were collected with 1.5, 1.6 and 1.7 mm aluminum shields; these are used in section~\ref{chain validation} as pile-up-free references for the calibration of the analysis chain. With a 2.2-mm-thick shield, nearly all acquired waveforms contained only a single pulse, as shown in figure~\ref{2.2 mm}. These data can be used to study the shape of the X-ray signal in USTC-IME LGADs, which will be discussed in section~\ref{global fit}.

\section{Data analysis}
\label{analysis}

Since the 2 ns repetition period of the SSRF is fixed by the radio-frequency structure of the storage ring and comparable to the response time of our detection system, pile-up effects occur between adjacent signal pulses in the DUT waveforms and are unavoidable in multi-pulse waveforms, as illustrated in figure~\ref{1 mm}. This phenomenon complicates the establishment of a stable baseline, thereby affecting the accurate determination of signal amplitudes and TOAs via CFD. In this section we first extracted the repetition period of the light source, using estimated TOAs. Then a global template fit was performed on the periodic multi-pulse waveforms to reduce dependence on pile-up, and a combined profile likelihood approach was used to derive the time resolution. Finally, control samples were utilized to estimate the systematic uncertainty and to correct the result.

\subsection{The relative TOA method}
\label{Preliminary analysis}
\label{cfd resolution}

\begin{figure}[htbp]
    \centering
    \begin{subfigure}[b]{0.5\textwidth}
        \includegraphics[width=\textwidth]{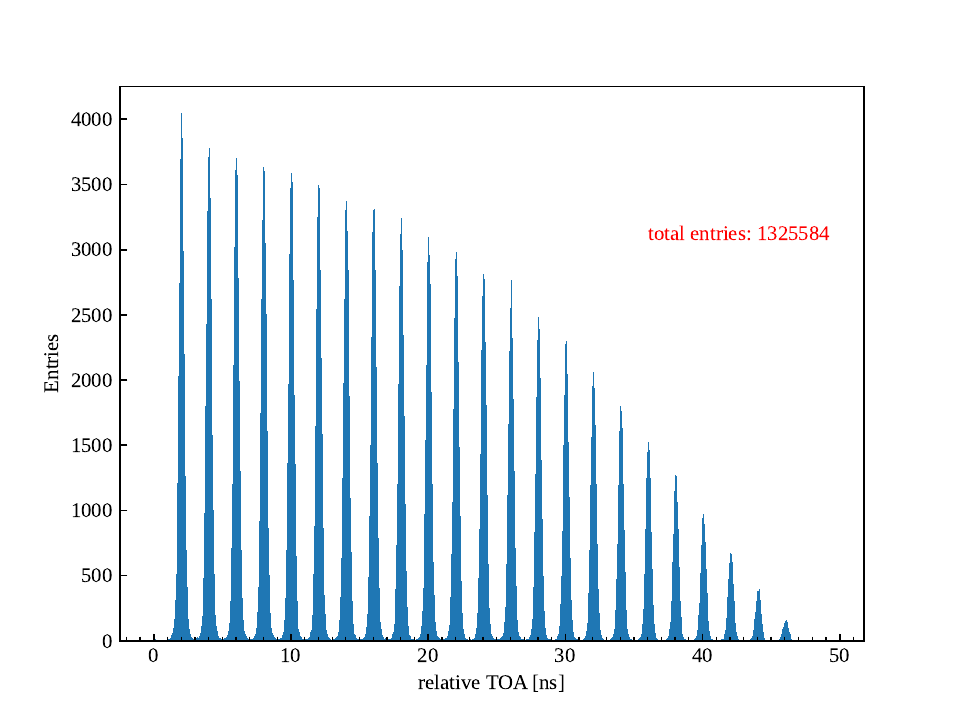}
        \caption{}
        \label{Distribution_of_delta_toa_all}
    \end{subfigure}
    \hfill
    \begin{subfigure}[b]{0.45\textwidth}
        \includegraphics[width=\textwidth]{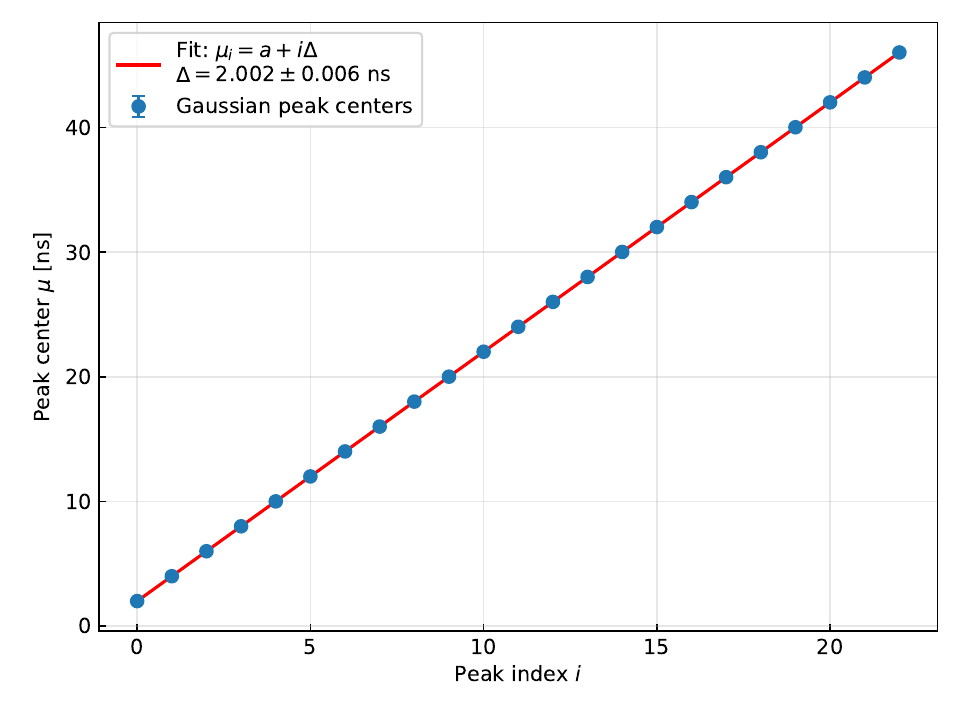}
        \caption{}
        \label{Linear_fit_of_peak_position}
    \end{subfigure}
    \caption{(a) Distribution of the relative TOA. (b) Fitted peak centers of figure~\ref{Distribution_of_delta_toa_all} and their linear fit result.}
    \label{fig:main}
\end{figure}

Because of the shielding effect of the DUT, the reference sensor predominantly captured single-pulse signals with stable baselines. Prior to data taking, waveforms of the DUT and the reference sensor were shifted to overlap on the oscilloscope. Therefore, the baseline of the reference sensor was adopted as an estimate for that of the DUT. For each coincidence event, the average voltage sample over the first 5 ns segment of the waveform of the reference sensor was selected as the estimated baseline. This estimated baseline was subsequently subtracted from the DUT waveform before determining the TOA. The TOA of each pulse, taken as the $50\%$ constant-fraction crossing of its leading edge, was computed with respect to that of the first pulse, giving the relative TOA. These waveforms were captured from the DUT within a time window of 50 ns, in which the consecutive signal pulses corresponded to successive beam spills. Thus, these relative TOAs represent the time intervals between beam spills. Figure~\ref{Distribution_of_delta_toa_all} shows the distribution of these relative TOAs, where only signal pulses with amplitudes exceeding 100 mV are considered, corresponding to the trigger threshold. The distribution exhibits multiple peaks, each of which was fitted with a Gaussian function to obtain its center. The resulting peak centers demonstrate good linearity, with a slope of 2.00 ns, as shown in figure~\ref{Linear_fit_of_peak_position}. This slope is consistent with the repetition period of the SSRF, indicating that LGADs are suitable for diagnostics of synchrotron radiation X-rays. However, these estimated relative TOAs are not suited for analyzing the time resolution at the sub-nanosecond level from pile-up waveforms, because the practical baseline of the DUT strongly depends on pile-up. Therefore, the global template fit described in the next subsection is used instead.

\subsection{Global template fit for fine time analysis}
\label{global fit}

In the previous study conducted at the SSRL, it was observed that the time resolution of LGADs is significantly worse when detecting X-rays compared to detecting MIPs~\cite{Xray2}. This degradation is primarily attributed to the large variation in the depth at which photons are absorbed. Unlike MIPs, which generate electron-hole pairs along their entire trajectories, incident photons are typically absorbed in a highly localized region. Consequently, the generated charge carriers must drift over different distances to reach the gain layer. The TCAD Sentaurus and Allpix Squared simulation described in section~\ref{simulation section} predicts that in a 50-\textmu\text{m}-thick LGAD sensor the maximum difference in drift time can exceed 500 ps, while the distribution has a root-mean-square (RMS) of 150.8 ps. This variation results in a fluctuating delay time of the signal. Therefore, this delay time is the key to understanding the time performance of LGADs in detecting X-rays.

Since all signal pulses originate from the same physical process---namely, the absorption of a photon followed by the drift of the released carriers---the expected pulse shape should be consistent, which can be characterized by a common signal template. This property enables a multi-peak fitting approach to extract the parameters of individual pulses, in which the delay time of each pulse is a parameter of the model, mitigating the pile-up effects. Such a methodology has been widely adopted in high energy physics, particularly for improving energy resolution~\cite{fit1,fit2}. In this work, we adapt this technique for fine time analysis. The fitting formula for waveform \(w\) is given by Equation \eqref{eq:1}:
\begin{equation}
\label{eq:1}
\begin{aligned}
y_{iw} = \sum_{k \in L_w} A_{kw} h(t_{iw} - \tau_{kw}) 
     &+ \sum_{j \in S_w} A_{jw} h(t_{iw} - \tau_{jw}) 
     + b_w(t_{iw})\,,
\\
\tau_{kw} = kT + \delta t_{kw} + \Delta t_w,\quad \tau_{jw} = jT + \delta t_{jw}& + \Delta t_w,\quad L_w,S_w\subseteq[-12, -11, \dots, 12, 13]\,,
\\
b_w(t_{iw}) = &\; a_w + c_w t_{iw}\,,
\end{aligned}
\end{equation}
where \(y_{iw}\) is the voltage sample at time \(t_{iw}\); \(L_w\) and \(S_w\) are the sets of serial numbers corresponding to large and small pulses according to the 100 mV trigger threshold; \(k\) and \(j\) are the indices determined by the relative positions of the respective pulses on the waveform; \(A_{kw}\) and \(A_{jw}\) are the amplitudes of the \(k\)-th and \(j\)-th pulses, respectively; \(h(t)\) is the normalized signal template; \(\tau_{kw}\) and \(\tau_{jw}\) are the reference times of the \(k\)-th and \(j\)-th pulses; \(T\) is the repetition period of the SSRF of 2 ns; \(\delta t_{kw}\) and \(\delta t_{jw}\) are the random delay times of interest; \(\Delta t_w\) is a waveform-level global time shift, from the fluctuation of the trigger signal with respect to the beam clock; and \(b_w(t)\) is a linear term accounting for potential noise.

The determination of \(\tau_{kw}\) and \(\tau_{jw}\) was the most critical step in the fitting procedure. In principle, they should be governed primarily by \(T\). However, given that there is no synchronous clock signal and the DUT was triggered by the reference sensor, the trigger time did not necessarily match the repetition period of 2 ns. Consequently, the waveform-level global time shift \(\Delta t_w\) must be incorporated into \(\tau_{kw}\) and \(\tau_{jw}\) to achieve proper alignment in time. In addition, the random delay time \(\delta t_{kw}\) and \(\delta t_{jw}\) must be taken into account. The waveform-level global time shift \(\Delta t_w\) is a nuisance parameter for determining $\delta t$. In this work, we take advantage of the waveforms with multiple pulses to eliminate the impact of this nuisance parameter. This technique can be applied to test setups with similar conditions.

\begin{figure}[htbp]
    \centering
    \begin{subfigure}[b]{0.45\textwidth}
        \includegraphics[width=\textwidth]{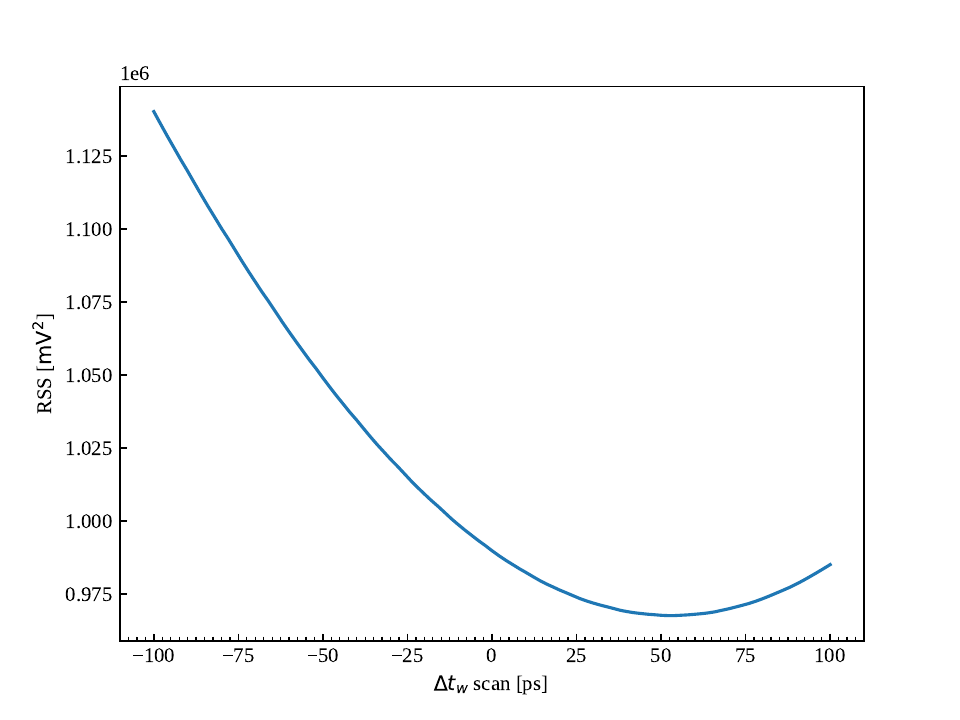}
        \caption{}
        \label{RSSDeltat1}
    \end{subfigure}
    \hfill
    \begin{subfigure}[b]{0.45\textwidth}
        \includegraphics[width=\textwidth]{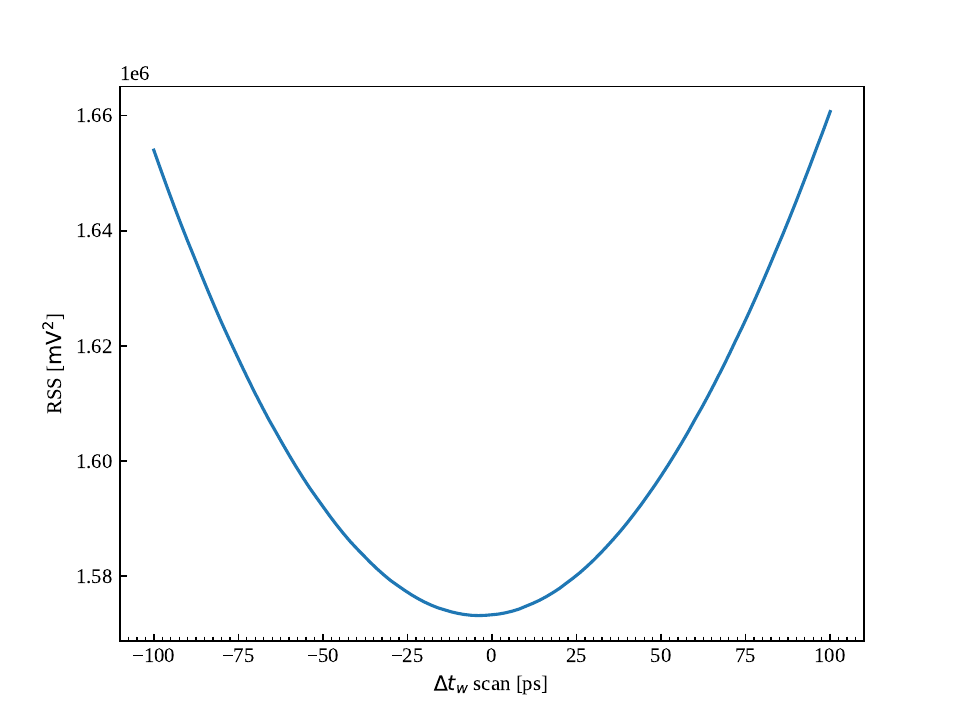}
        \caption{}
        \label{RSSDeltat2}
    \end{subfigure}
    \caption{Effect of the \(\Delta t_w\) scan on two representative waveforms. The \(\Delta t_w\) was selected to minimize the RSS.}
    \label{per beam spill}
\end{figure}

Fitting \(\delta t_{kw}\) and \(\delta t_{jw}\) for every pulse individually, together with \(\Delta t_w\), would introduce an excessively large number of free parameters and increase the complexity of the fit. To address this issue, a simplified treatment was adopted. \(\Delta t_w\) was not treated as a floating parameter in the fit; instead, it was determined by scanning within a range of $\pm0.1$ ns around the TOA of the pulse closest to the trigger time, with a step size of 2 ps, so as to minimize the residual sum of squares (RSS). The TOA was shifted into the range of $-1$ ns to 1 ns after taking modulo 2 ns, thereby ensuring that the time shift does not exceed $T/2+0.1$ ns. Figure~\ref{per beam spill} illustrates the effect of the \(\Delta t_w\) scan. Furthermore, given that the fitting accuracy for small pulses is significantly worse than that for large ones, these two groups were optimized separately. For small pulses and regions without signal pulses, \(\delta t_{jw}\) was not considered, and \(\tau_{jw}\) was determined solely from \(T\) and \(\Delta t_w\). In contrast, for large pulses, \(\tau_{kw}\) was constructed using only \(T\) and the fitted \(\delta t_{kw}\), without explicitly incorporating \(\Delta t_w\) into the expression. This was to prevent the known \(\Delta t_w\) from dominating the RSS minimization and thereby degrading the fit quality. It should be noted that \(\delta t_{kw}\) obtained in this manner inherently includes \(\Delta t_w\), and thus the spread taken directly from the \(\delta t_{kw}\) distribution also incorporates the contribution from waveform-to-waveform variations in the time shift. This uncertainty ultimately originates from the lack of a synchronous clock signal. To facilitate the fitting, linear approximation in terms of \(\delta t_{kw}\) given by Equation \eqref{eq:2} 
\begin{equation}
\label{eq:2}
y_{iw} \approx \sum_{k \in L_w} A_{kw} [h(t_{iw} - kT) - \dot{h}(t_{iw} - kT)\delta t_{kw}] 
     + \sum_{j \in S_w} A_{jw} h(t_{iw} - jT - \Delta t_w) 
     + a_w + c_w t_{iw}\,,
\end{equation}
was applied for the signal template. The accuracy of this approximation was checked using the Gauss--Newton minimization method with the original template function. The difference in the obtained time resolution with respect to the approximated linear fit is  $0.5 \pm 0.6$ ps, so the linear approximation is adopted.

\begin{figure}[htbp]
\centering
\includegraphics[width=.6\textwidth]{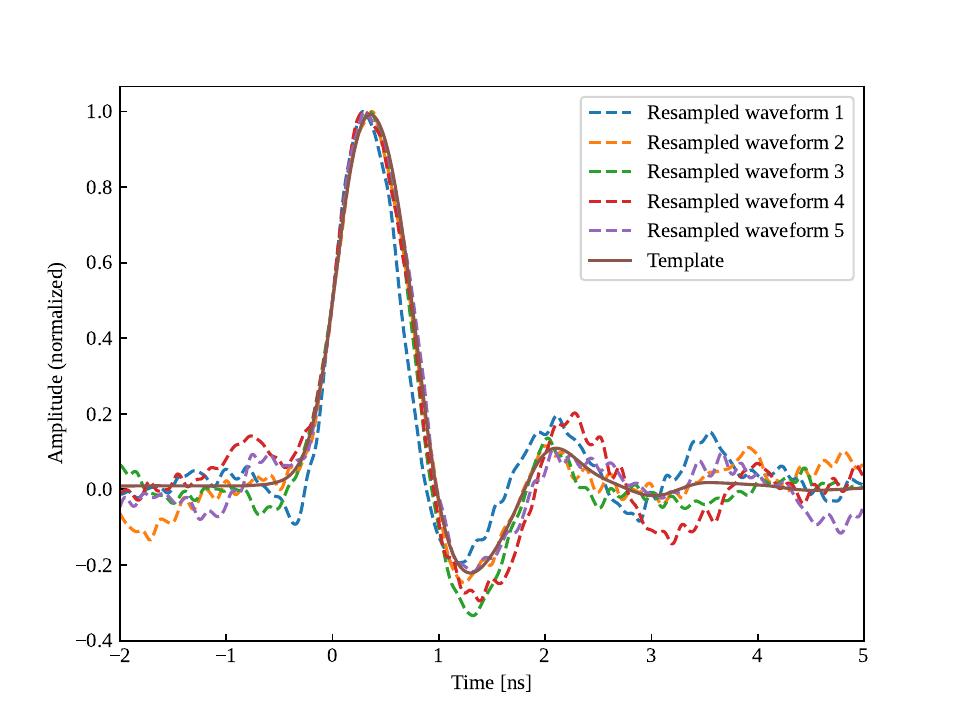}
\caption{Signal template and some resampled waveforms.\label{template}}
\end{figure}

\begin{figure}[htbp]
\centering
\includegraphics[width=.6\textwidth]{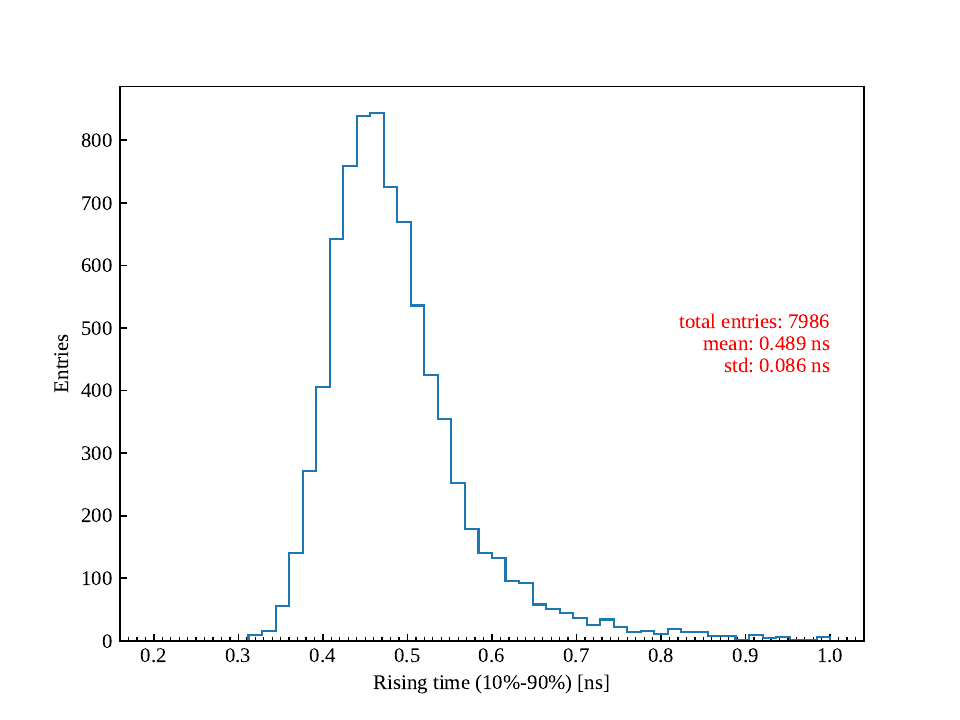}
\caption{Distribution of the rising time (from 10\% to 90\% of the rising edge) of pulses used to obtain the signal template.\label{The distribution of the rising time}}
\end{figure}

The signal template was derived from a set of waveforms acquired with a 2.2-mm-thick aluminum shield, as described in section~\ref{Experimental setup}. In these single-pulse waveforms the baseline outside the main pulse shows a small residual structure, originating from the ringing of the fast preamplifier and of the oscilloscope front end together with cable reflections. Since the fitting model represents the baseline by only a constant and a linear term, this residual structure is not fully absorbed by the model. However, its effect on the extracted timing is negligible because its RMS amplitude is below 10 mV in most waveforms, which is much smaller than the 100 mV trigger threshold. This residual structure is therefore expected to have a negligible effect on TOA determination. Moreover, the average waveform was adopted as the final template, so the ringing was further suppressed. After excluding waveforms characterized by excessive noise or containing more than one pulse, a total of 7986 waveforms were retained for analysis. The construction procedure was as follows: first, the 50\% CFD TOA of each pulse was shifted to $t = 0$ for temporal alignment; then, an interpolation was conducted on a unified time grid spanning from $-2$ ns to 5 ns with a step of 0.01 ns to cover the entire pulse; and the amplitudes of these pulses were normalized to unity; finally, the template was obtained by averaging the normalized pulses, as shown in figure~\ref{template}. The common-template assumption is supported by the narrow spread of the pulse shapes: the 10\%--90\% rising time has an RMS of 86 ps around a mean of about 0.49 ns (figure~\ref{The distribution of the rising time}), and the amplitude normalization absorbs the pulse-to-pulse amplitude variation. The photon absorption depth affects the drift time and therefore mainly shifts the onset of the signal in time. Since the induced signal is dominated by the charge carriers crossing the gain layer, where the drift field is the highest, the drift through the bulk contributes little to the waveform shape, so a fixed template can be used. The residual shape variation associated with the absorption depth is negligible for the mean delay, because it mainly displaces the onset of the signal; any pulse-to-pulse component of the shape variation enters the spread and is included in the empirical calibration of section~\ref{chain validation}.

\begin{figure}[htbp]
\centering
\includegraphics[width=.6\textwidth]{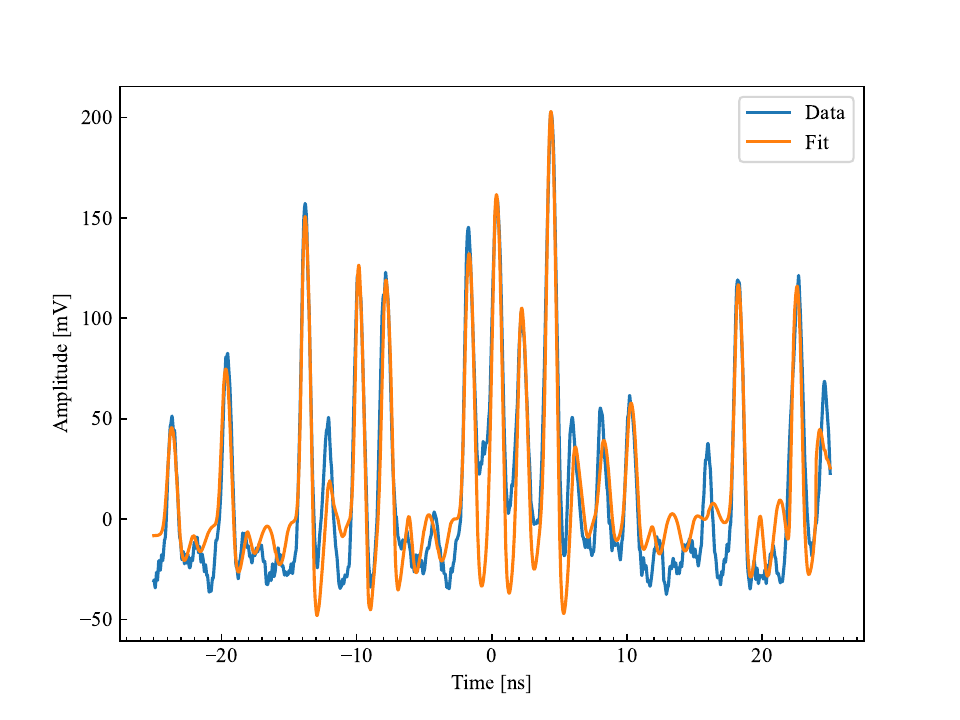}
\caption{Representative fitting result.\label{fitted waveform}}
\end{figure}

\begin{figure}[htbp]
\centering
\includegraphics[width=.6\textwidth]{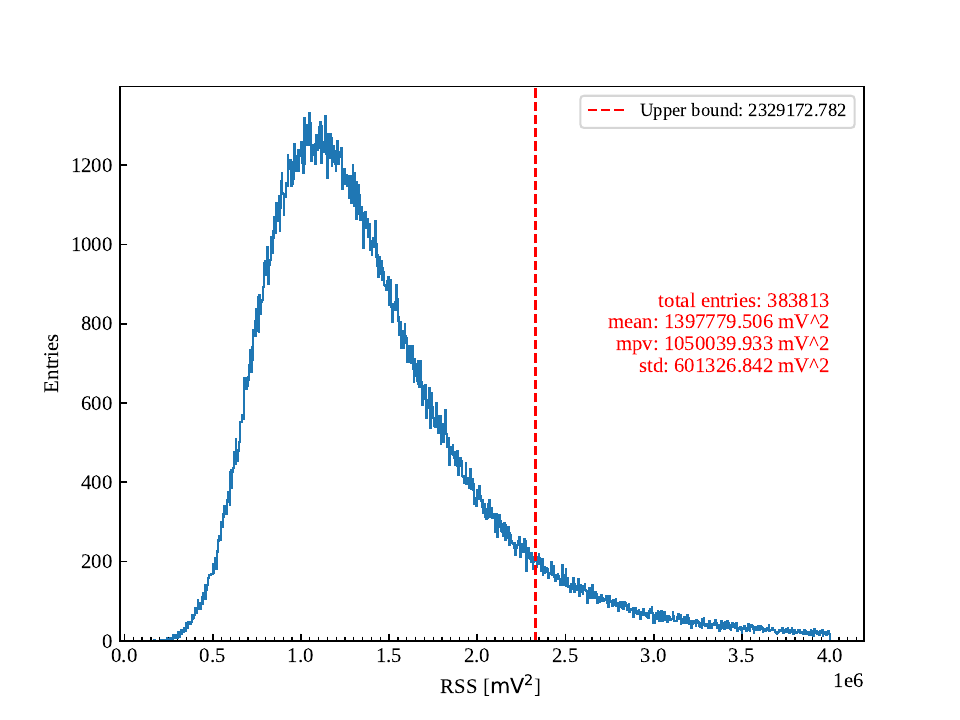}
\caption{Distribution of the RSS. The cut is determined by the median plus 1.5 times the interquartile range.\label{RSS}}
\end{figure}

\begin{figure}[htbp]
\centering
\includegraphics[width=.6\textwidth]{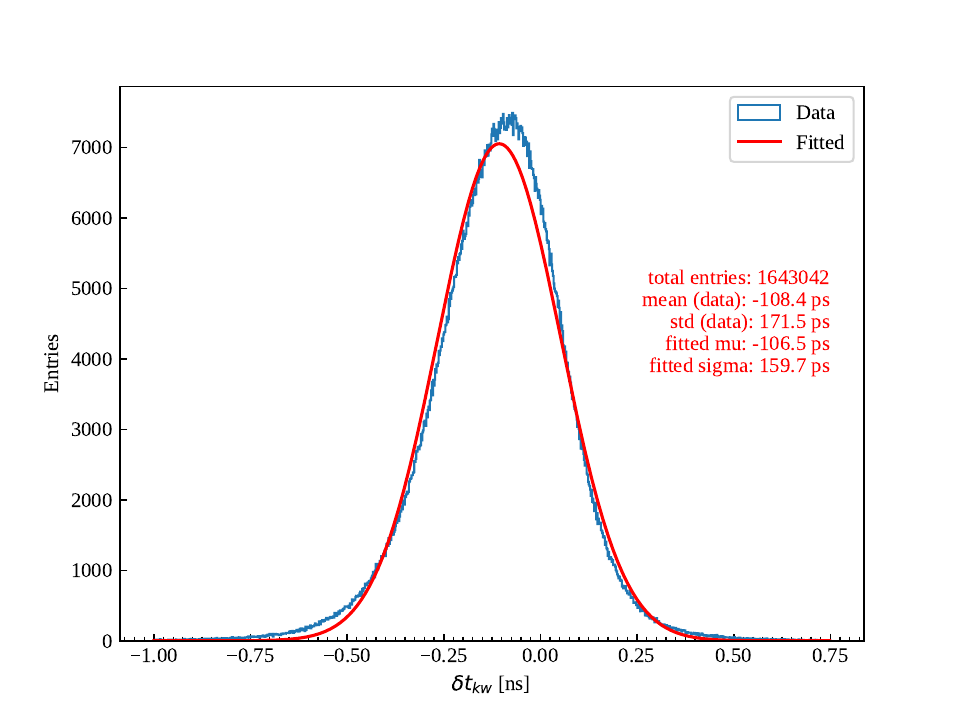}
\caption{Distribution and Gaussian fit of \(\delta t_{kw}\).\label{delta_tk}}
\end{figure}

Figure~\ref{fitted waveform} presents a representative fitting result, showing that the large pulses are well characterized. After rejecting waveforms with excessive RSS as shown in figure~\ref{RSS} and excluding reconstructed pulses with an amplitude below 100 mV or an excessively large undershoot, the distribution of the fitted \(\delta t_{kw}\) values is presented in figure~\ref{delta_tk}. A direct Gaussian fit turns out to be suboptimal, with the fitted mean deviating from zero. This is primarily attributed to the waveform-to-waveform variations in \(\Delta t_w\). This effect is removed as follows.

The quantity \(\delta t_{kw}\) should follow a certain distribution, whose RMS is the time resolution of the LGAD in detecting X-rays. 
In the approach above, the mean values of \(\delta t_{kw}\) in waveform \(w\) are shifted by the influence of \(\Delta t_w\), but the delays of all pulses, in every waveform, should share a common RMS, denoted by \(\sigma\) hereafter. A combined profile likelihood approach was employed to estimate the common time resolution \(\sigma\)~\cite{Pawitan}, after the data cleaning described above. For each waveform $w$, the mean $\mu_w$ was treated as a nuisance parameter and profiled out by substituting its maximum likelihood estimate $\hat{\mu}_w = \bar{\delta t}_w$, where \(\bar{\delta t}_w\) is the mean of \(\delta t_{kw}\) within waveform $w$. The resulting profiled negative log-likelihood for each waveform is:
\begin{equation}
\textrm{NLL}_w(\sigma) = n_w \ln(\sigma) + \frac{1}{2\sigma^2}\sum_{k \in L_w}(\delta t_{kw} - \bar{\delta t}_w)^2
\end{equation}
where $n_w$ is the number of large pulses in waveform $w$. The combined profile likelihood was obtained by summing over all waveforms:
\begin{equation}
\textrm{NLL}_\textrm{total}(\sigma) = \sum_{w=1}^{W} \textrm{NLL}_w(\sigma),
\end{equation}
where \(W\) is the number of waveforms. The maximum-likelihood estimator (MLE) is:
\begin{equation}
\hat{\sigma}^2 = S^2/N,\quad S^2 = \sum_{w=1}^{W}\sum_{k \in L_w}\bigl(\delta t_{kw} - \bar{\delta t}_w\bigr)^2,\quad N = \sum_w n_w.
\end{equation}
Since profiling $\mu_w$ consumes exactly one degree of
freedom per waveform, \(S^2/\sigma^2\) follows the \(\chi^2\) distribution with \(N-W\) degrees of freedom.
Therefore, $\hat{\sigma}^2$ has expectation:
\begin{equation}
E[\hat{\sigma}^2] = \frac{(N-W)}{N}\sigma^2,
\label{eq:bias}
\end{equation}
that is, $\hat{\sigma}^2$ underestimates $\sigma^2$ by the factor $\frac{N-W}{N}$. To remove this bias, the restricted maximum-likelihood (REML) estimator~\cite{REML}
\begin{equation}
\tilde{\sigma}^2 = \frac{S^2}{N-W},
\label{eq:REML}
\end{equation}
is used as the central value, and the 68\% confidence interval is derived from the $\chi^2_{N-W}$ distribution,
\begin{equation}
\sigma \in \left[\sqrt{\frac{S^2}{\chi^2_{N-W,\,1-\alpha/2}}},\;
\sqrt{\frac{S^2}{\chi^2_{N-W,\,\alpha/2}}}\right], \quad \alpha = 0.32,
\label{eq:chi2CI}
\end{equation}
where $\chi^2_{N-W,\,p}$ denotes the $p$-quantile.

\begin{figure}[htbp]
\centering
\includegraphics[width=.6\textwidth]{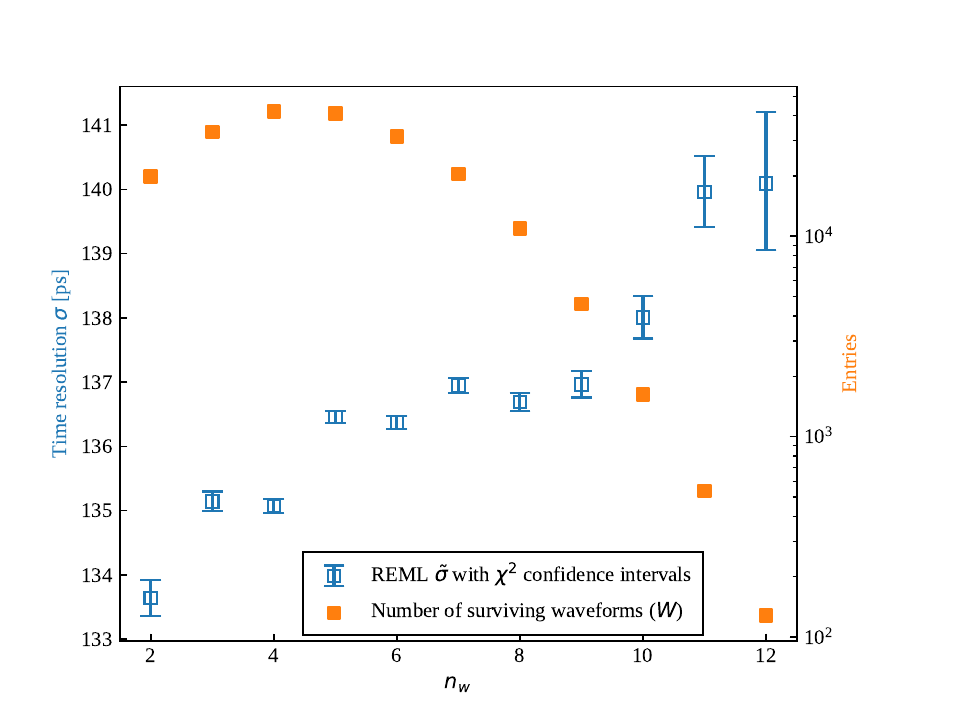}
\caption{The time resolution and the number of surviving waveforms as a function of the number of pulses in the waveform; for a given $n_w$, the number of pulses entering the estimation is $n_w \times W$. Open squares: the REML estimate $\tilde{\sigma}$ of the global template fit with uncertainties; filled markers: the number of surviving waveforms $W$ (right axis).\label{delta_tk_sigma}}
\end{figure}

To assess the dependence of $\tilde{\sigma}$ on pile-up, the number of large pulses in a waveform was then scanned. Figure~\ref{delta_tk_sigma} presents, as a function of $n_w$, the REML estimate $\tilde{\sigma}$ of the global template fit together with the number of surviving waveforms. $\tilde{\sigma}$ increases from $133.6$ ps at $n_w=2$ to $140.1$ ps at $n_w=12$, and is nearly independent of $n_w$ over the range $5 \le n_w \le 9$: the five estimates in this interval are consistent. The residual dependence of $\tilde{\sigma}$ on $n_w$ is attributed to the beam intensity: waveforms with many (few) large pulses come from periods of higher (lower) intensity, where the pile-up is stronger (weaker). The $n_w$-independent range $5 \le n_w \le 9$ was chosen and it gives $\tilde{\sigma} = 136.6 \pm 0.1$ ps, obtained with $\bar{n}_w = 6.14$. Over the range $2 \le n_w \le 10$, within which the statistical uncertainty remains below $\pm 1.0$ ps, $\tilde{\sigma}$ spans 133.6-138.0 ps, giving a systematic uncertainty of $\pm 2.2$ ps from the level of pile-up.

\subsection{Calibration of the analysis chain on control samples}
\label{chain validation}

\begin{figure}[htbp]
\centering
\includegraphics[width=.6\textwidth]{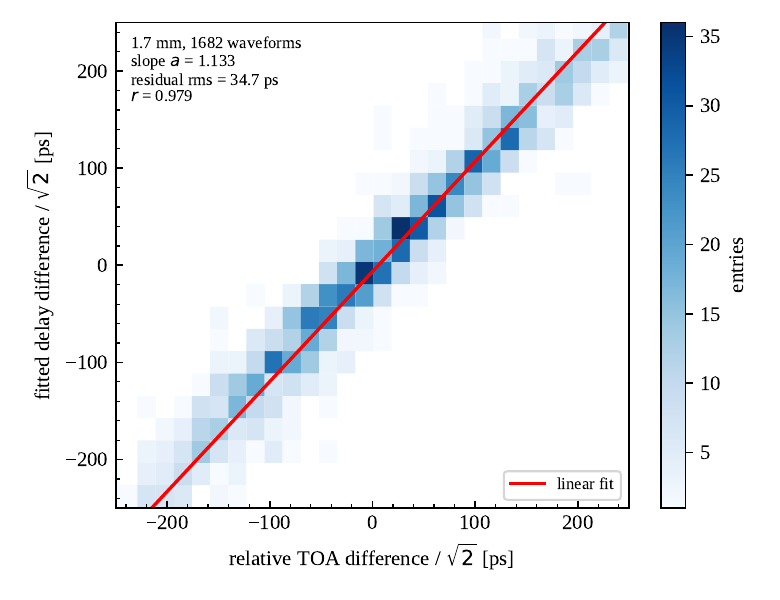}
\caption{Relation between the relative TOA difference and the fitted delay difference between the two pulses, waveform by waveform, for the 1.7 mm sample; both axes are the per-pulse difference divided by $\sqrt 2$, and the line is the result of a linear regression.\label{calibration relation}}
\end{figure}

The whole analysis chain should be calibrated on waveforms free of pile-up, so we selected waveforms containing exactly two pulses above 100 mV, separated by 5 to 50 ns, and no other structure above 50 mV. Only 35 waveforms of the primary data set satisfy this requirement, whereas the three control samples described in section~\ref{Experimental setup} contain enough of them to perform the calibration. There the pulses are separated and the baseline is well defined, so the limitation discussed in section~\ref{cfd resolution} does not apply. The relative TOA method is therefore a reliable reference: in the absence of pile-up, the TOAs of two pulses within a waveform are measured independently, so the RMS of their difference---with the distribution folded modulo the 2 ns period---divided by $\sqrt 2$ gives the time resolution of the DUT for a single pulse. This quantity, denoted by ${\rm RMS}_{\rm CFD}$, is an estimator of the time resolution, similar to the REML estimator. For comparison, both estimators were evaluated on the same waveforms. The two estimates are found to be tightly correlated waveform by waveform. Figure~\ref{calibration relation} illustrates their linear correspondence for the 1.7 mm sample. Table~\ref{tab:calibration} summarizes the time resolutions obtained with the two methods and their ratios for the three control samples. The estimated resolutions differ among the samples. The ratios between the two estimators are used to derive the correction factor to calibrate the REML method. The global template fit yields a worse time resolution by approximately $13$--$16\%$. Applying these ratios to the measured value of $136.6$ ps yields a corrected mean of $119.7$ ps. Half of the spread among the three ratios, $\pm 1.4$ ps, is assigned as the uncertainty associated with this correction.

\begin{table}[htbp]
\centering
\small
\setlength{\tabcolsep}{4pt}
\renewcommand{\arraystretch}{1.5}
\begin{tabular}{@{} c c c c @{}}
\hline
Aluminum shield [mm] & Relative TOA method [ps] & Global template fit [ps] & Ratio \\
\hline
1.5 & $138.8\pm 4.0$ & $157.0^{+4.7}_{-4.4}$ & $1.131\pm0.033$ \\
1.6 & $146.7\pm 2.9$ & $166.6^{+3.4}_{-3.2}$ & $1.136\pm0.022$ \\
1.7 & $146.7\pm 2.5$ & $169.9^{+3.0}_{-2.9}$ & $1.158\pm0.020$ \\
\hline
\end{tabular}
\caption{Time resolution obtained with the two methods on the pile-up-free waveforms of the three control samples. The two methods are evaluated on the same waveforms and the uncertainties are statistical.}
\label{tab:calibration}
\end{table}

The time resolution of the LGAD obtained with the global template fit, together with its uncertainties, corrected as described above, is

\begin{equation}
\label{eq:final}
\sigma_{\rm corr} = 119.7 \pm 0.1\,(\text{stat.}) \pm 2.6\,(\text{syst.})~\text{ps}\,,
\end{equation}

\noindent where the statistical uncertainty is that of the measured spread, and the systematic uncertainty combines the dispersion of the correction factors obtained from the three control samples with that of the pile-up level in section~\ref{global fit}. Table~\ref{tab:systematics} summarizes all component of the uncertainties.

\begin{table}[htbp]
\centering
\small
\setlength{\tabcolsep}{3pt}
\begin{tabular}{@{} l c p{5.6cm} l @{}}
\hline
Source of uncertainty & Estimate [ps] & Method & Type \\
\hline
Statistical (measured spread)        & $0.1$ & profile likelihood (section~\ref{global fit}) & stat. \\
Correction of the analysis chain      & $1.4$  & three control samples (table~\ref{tab:calibration}) & syst. \\
Dependence on the pile-up level ($n_w$ window)            & $2.2$  & scan of the bins (section~\ref{global fit}) & syst. \\
\hline
Total (statistical)                   & $0.1$ & & \\
Total (systematic)                   & $2.6$ & & \\
\hline
\end{tabular}
\caption{Statistical and systematic uncertainties of the corrected time resolution.}
\label{tab:systematics}
\end{table}

\section{Simulation for a better understanding}
\label{simulation section}

To better understand the degradation in time resolution, we carried out simulations using TCAD Sentaurus and Allpix Squared. TCAD Sentaurus was employed to compute the electric field distribution within the LGAD sensor with a bias voltage of 180 V, following the fabrication process described in~\cite{XYANG}. Allpix Squared was utilized to simulate photon absorption and charge carrier transport within the LGAD sensor. In the simulation, the active region consists of a 50-\textmu\text{m}-thick silicon layer, topped with a 3-\textmu\text{m}-thick silicon dioxide layer. On the backside, a 300-\textmu\text{m}-thick silicon substrate and a 1-\textmu\text{m}-thick aluminum electrode are included. The photon source is configured as a single-photon beam with a fixed energy of 10 keV, focused to a spot size of approximately 500 \textmu\text{m} in diameter, pointing to the surface of the sensor. The simulation is performed at a temperature of 28~$\tccentigrade$ in an air environment. The simulation settings are consistent with the experimental conditions. Charge carrier multiplication is intentionally disabled, as this study focuses on the charge transport dynamics prior to carriers reaching the gain layer.

\begin{figure}[htbp]
    \centering
    \begin{subfigure}[b]{0.45\textwidth}
        \includegraphics[width=\textwidth]{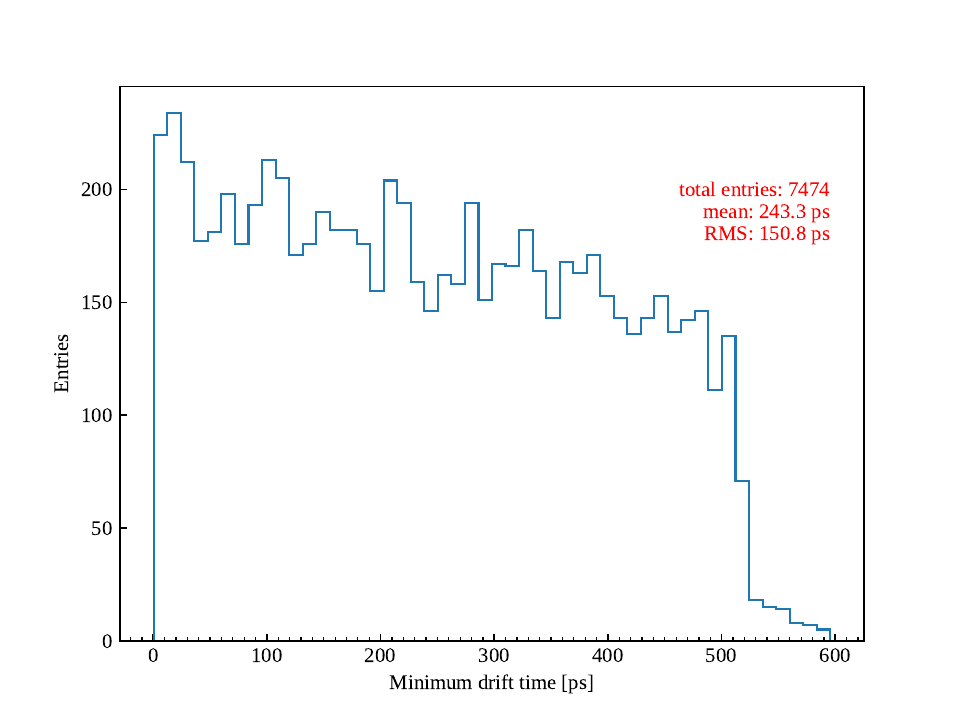}
        \caption{}
        \label{drift time}
    \end{subfigure}
    \hfill
    \begin{subfigure}[b]{0.45\textwidth}
        \includegraphics[width=\textwidth]{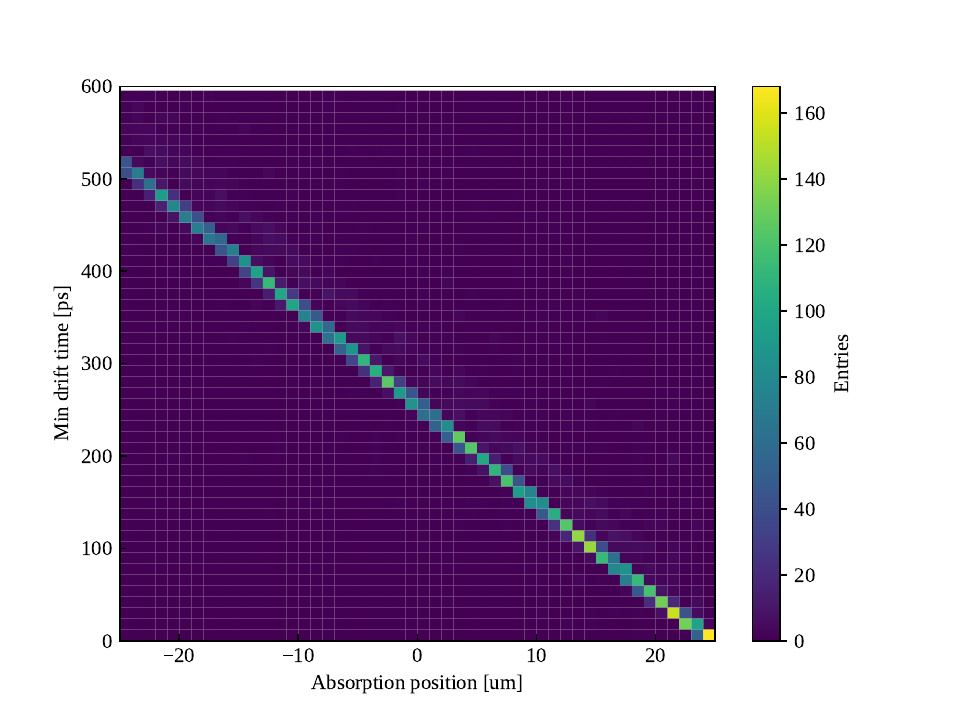}
        \caption{}
        \label{drift time and absorption depth}
    \end{subfigure}
    \caption{(a) Distribution of the minimum drift time of electrons in the simulation. (b) Correlation between the minimum drift time and the absorption position of the photons obtained from simulation. The absorption position of 25~\textmu m means the photons are absorbed on the upper boundary of the active region.}
    \label{simulation}
\end{figure}

The minimum time required for generated electrons to reach the gain layer was studied with respect to the injection time, which corresponds to the delay of the signal onset. Figure~\ref{drift time} presents the distribution of this delay, with an RMS of 150.8 ps. Figure~\ref{drift time and absorption depth} illustrates the correlation between the minimum drift time and the absorption position of the photons. The simulation shows that variations in the photon absorption depth, and hence in the electron drift time, dominate the time resolution. The difference from the real experiment is expected: the simulation describes the sensor alone, without the aluminum shield in front of it, and it does not include the effects of charge multiplication. It also considers photons absorbed at all depths, whereas in the measurement signals corresponding to an incomplete multiplication can be undetected because of the trigger threshold. The simulated spread therefore represents the largest value expected for the detector, and, due to this selection, the spread of the observed drift time should be lower than that in the simulation.

\section{Conclusions}
\label{Conclusions}

The time performance of a USTC-IME version 2.1 LGAD under 10 keV focused synchrotron X-rays was studied at the SSRF. The tested device successfully resolved the 2 ns repetition period of the SSRF, indicating that LGADs can be used for diagnostics of synchrotron radiation X-rays. Previous studies have shown that the time resolution of LGADs is significantly worse when detecting X-rays compared to detecting MIPs. This has been explained by the large variation of the drift lengths of the charge carriers. In this work, a global template fit was employed to extract the random delay time of each individual signal pulse from the multi-pulse waveforms; it does not rely on a baseline distorted by pile-up, and the extracted delay depends only weakly on the number of pulses in the waveform. A combined profile likelihood approach, with a restricted maximum-likelihood correction for the per-waveform degrees of freedom, was then utilized to estimate the \(\sigma\) of this delay time. This method takes advantage of the multiple pulses on a waveform to profile out the nuisance parameter associated with the shift of the trigger signal with respect to the DUT readout. The obtained time resolution of the LGAD is $119.7 \pm 0.1$ (stat.) $\pm 2.6$ (syst.) ps. As expected, this is significantly worse than the 39.6 ps obtained for MIP detection. The chain was calibrated on control samples, where the relative TOA method was used as a reference. The two timing methods were compared on the same waveforms, and the corresponding correction was applied to the result. A simulation using TCAD Sentaurus and Allpix Squared was performed to study the process by which X-rays generate signals in LGADs. It shows that the variation in the photon absorption depth is the dominant contribution to the spread of the timing response.

\acknowledgments

This work is partially supported by National Natural Science Foundation of China (grant No. 12535012 and No. W2443005). The authors would like to thank staff of the BL16B1 beamline of the Shanghai Synchrotron Radiation Facility (SSRF) for their experimental support.





\end{document}